\documentclass[conference]{IEEEtran}
\IEEEoverridecommandlockouts
\usepackage{cite}
\usepackage{amsmath,amssymb,amsfonts}
\usepackage{amsthm}
\usepackage{algorithmic}
\usepackage{graphicx}
\usepackage{textcomp}
\usepackage{xcolor}
\usepackage{braket}
\usepackage[lined, boxed,ruled,vlined]{algorithm2e}
\def\BibTeX{{\rm B\kern-.05em{\sc i\kern-.025em b}\kern-.08em
    T\kern-.1667em\lower.7ex\hbox{E}\kern-.125emX}}
    
\newtheorem{definition}{Definition}

\newcommand{\etal}{\emph{et al.\ }}

\begin{document}

\title{A Depth-Aware Swap Insertion Scheme \\ for the Qubit Mapping Problem\\
}

\author{
    \IEEEauthorblockN{Chi Zhang\IEEEauthorrefmark{1}  Yanhao Chen\IEEEauthorrefmark{2}  Yuwei Jin\IEEEauthorrefmark{2} \\ Wonsun Ahn\IEEEauthorrefmark{1}  Youtao Zhang\IEEEauthorrefmark{1}  Eddy Z. Zhang\IEEEauthorrefmark{2}}
    \vspace{1.5ex}
    \IEEEauthorblockA{\IEEEauthorrefmark{1}University of Pittsburgh
    \\\{chz54, wahn\}@pitt.edu, zhangyt@cs.pitt.edu}
    \vspace{1.5ex}
    \IEEEauthorblockA{\IEEEauthorrefmark{2}Rutgers University
    \\chenyh64@gmail.com, yj243@scarletmail.rutgers.edu, eddy.zhengzhang@gmail.com}
}

\maketitle

\begin{abstract}
The rapid progress of physical implementation of quantum computers paved the way
of realising the design of tools to help users write quantum programs for any
given quantum devices. The  physical constraints inherent to the current NISQ
architectures prevent most quantum algorithms from being directly executed on 
quantum devices. To enable two-qubit gates in the algorithm, existing works
focus on inserting SWAP gates to dynamically remap logical qubits to physical
qubits. However, their schemes lack the consideration of the depth of generated
quantum circuits. In this work, we propose a depth-aware SWAP insertion scheme
for qubit mapping problem in the NISQ era. 
\end{abstract}

\section{Introduction}

Quantum computing has exhibited its theoretical advantage over classical
computing by showing impressive speedup on applications including large integer factoring
\cite{Shor:FOCS94}, database search \cite{grover+:stoc96},
and quantum simulation \cite{peruzzo+:nature14}. It is
considered to be a new computational model that may have a subversive impact on
the future, and has attracted major interests of a large number of researchers
and companies. 

With the advent of advanced manufacturing technology, the industry is able to build
small-scale quantum computers -- Noisy Intermediate-Scale
Quantum~\cite{Preskill:Quantum18} (NISQ) devices. A NISQ device is equipped with
dozens to hundreds of qubits. IBM~\cite{Knight:technologyreview17} released its
53-qubit quantum computer in October 2019 and has made it available for
commercial use. Google~\cite{Kelly:googleai2018} released the 72-qubit
\emph{Bristlecone} quantum computer in March 2018. Other companies including Intel
\cite{Hsu:CES18}, Rigetti \cite{Rigetti}, and IonQ, have released their
quantum computing devices with dozens of qubits. 
The current NISQ technology may not be
perfect, but it's a good first step towards the more powerful quantum devices in the future.

In order to map high level quantum programs to NISQ devices, it is important to
overcome two obstacles. First, to be able to execute a quantum circuit, it is necessary to map logical qubits to physical
qubits with respect to architecture and program coupling constraints. Any quantum program
can be implemented using an universal gate set \cite{Nielsen+:2002book} of a small number
of elementary gates. For
instance, the \emph{\{H, CNOT, S, T\} } set is an universal gate set, in which
the \emph{\{H, S, T\} } gates are single qubit gates, the \emph{CNOT} gate is a
two-qubit gate. The two-qubit gate must be mapped to two qubits that are
physically connected. However, in real quantum architecture, qubits may have
limited connection and not every two qubits are connected, as shown in the IBM
\emph{QX2} architecture in Fig. \ref{fig:hardware} (a). For this reason, a quantum
circuit is not directly executable on a NISQ device, unless circuit
transformation is performed. The common practice is to insert \emph{SWAP}
operations to dynamically remap the logical qubit such that the
transformed circuit is hardware-compliant for each (set of) two-qubit gate(s).

Second, it is critical that the depth of a quantum circuit be minimized for the
NISQ device. A qubit is volatile and error prone. It gradually decays over
time and may have phase and bit flip errors. It may completely lose
its state after a certain period of time, called
\emph{coherence time}. Quantum error correction (QEC) codes can
detect error syndromes and fix them. However, QEC needs to use a large
number of redundant physical qubits. A realistic QEC circuit may need more than
10,000 physical qubits, which is not possible for today's NISQ device. Without
QEC, a program must terminate within a threshold amount of time. The depth of the
circuit, which is the number of steps the circuit executes, must be optimized.
IBM proposed the metric of \emph{quantum volume} \cite{Cross_2019} for evaluating the
effectiveness of quantum computers which accounts for not only the width of the
circuit (the number of qubits), but also the depth, how many steps the circuit
can execute.

Transforming the logical circuit into a hardware-compliant one will inevitably result in
increased gate count and circuit depth. Most previous work for qubit mapping
\cite{Li+:ASPLOS19, Wille+:DAC19, Zulehner+:ICRC17, Zulehner+:DATE18, IBMQiskit, Siraichi+:CGO18} focus on 
minimizing the number of inserted gates, but not the depth of the
transformed circuit. However, even if the gate count is small, it does not
necessarily mean the depth of the circuit is small, due to the dependence
between different gates. We discover that previous work that aims to minimize number of
inserted gate may significantly increase the depth of the circuit (in
Section \ref{sec:eval}). For instance, the Sabre approach by Li \etal
\cite{Li+:ASPLOS19} reduces the gate count by 1.1\%, but increases the depth of the
10-qubit \emph{QFT} circuit by over 44.5\%. The two studies \cite{tannu+:asplos19,murali+:asplos19} stress the
importance of taking into consideration the variability in the qubit (link)
error rates, but they do not directly address the issue of the increased circuit depth. 

The depth of the circuit, as mentioned above, is critical and determines if a quantum program
is executable on a NISQ device with respect to its physical limits. 
In this paper, we propose the first {depth-aware qubit mapping} scheme for
quantum circuits running on arbitrary qubit connectivity hardware. Our
depth-aware qubit mapper searches for the mapping that minimizes the
transformed circuit depth and keeps the gate count within a
reasonable range. Our results show we can reduce the depth of the transformed circuit by
up to 30\% compared with two best known qubit mappers \cite{Li+:ASPLOS19,
Wille+:DAC19}, and in the meantime, have on average less than 3\% additional
gates over a large set of representative benchmarks.

\section{Background and Motivation}
\vspace{-1pt}

\subsection{Quantum Computing Basics}

\subsubsection{Qubit}
A quantum bit or qubit, is the counterpart to classical bit in the realm of
quantum computing. Different from a classical bit that represents either `1' or
`0', a qubit is in the coherent superposition of both states. It is considered as a two-state quantum system that exhibits the peculiarity of quantum mechanics~\cite{Nielsen+:2002book}. An example is the spin of the electron that the two states can be spin up and spin down.  


\subsubsection{Quantum Gates}
There are two types of basic quantum gates. One type of basic gates is the
single-qubit gate, a unitary quantum operation that can be abstracted as the
rotation around the axis of the Bloch sphere\cite{Nielsen+:2002book} which
represents the state space of one qubit. A single qubit-gate can be
parameterized using two rotation angles around the axes. There are several
elementary single-qubit gates including the Hadamard (H) gate, the phase (S)
gate, and the $\pi/8$ (T) gate \cite{Nielsen+:2002book}. The other
type of basic gates is the multi-qubit gate. However, all complex quantum gates can
be decomposed into a sequence of single qubit gates H, S, T, and the two-qubit CNOT
gate. Thus we only focus on the two-qubit CNOT gate. The CNOT gate operates on
two qubits which are distinguished as a control qubit and a target qubit. If the
control qubit is 1, the CNOT gate flips the state of the target qubit, otherwise,
the target qubit remains the same. 

\subsubsection{Quantum Circuit}
Quantum circuit is composed of a set of qubits and a sequence of quantum
operations on these qubits. There are various ways to describe the quantum
circuits. One way is to use the quantum assembly language called
OpenQASM~\cite{Andrew+:arxiv17} released by IBM. Another way is to use the
circuit diagram, in which qubits are represented as horizontal lines and quantum
operations are the different blocks on those lines. In Fig. \ref{fig:motivation}
(a), we show a simple example of quantum circuit diagram. A single-qubit gate is denoted
as a square on the line, and one CNOT gate is represented by a line connecting
two qubits and a circle enclosing a plus sign.


\subsection{Qubit Mapping and Depth-Awareness}
\begin{figure}[!ht]
  \centering
  \includegraphics[width=0.8\linewidth]{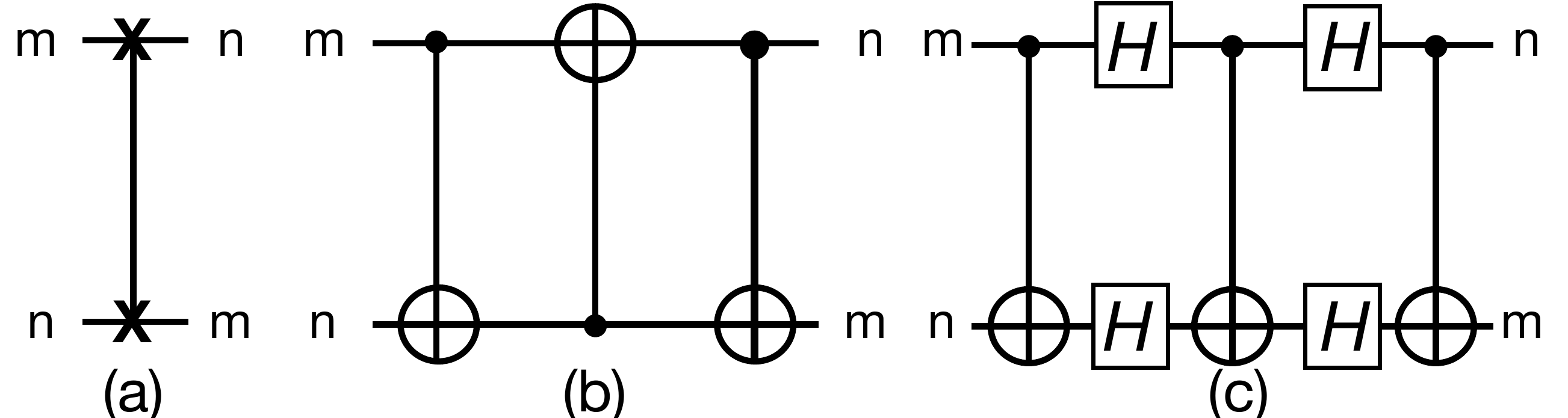}
  \caption{Implementation of a SWAP operation}
  \label{fig:swapgate}
  \vspace{-10pt}
\end{figure}

To enable the execution of a quantum circuit, the
logical qubits in the circuit must be mapped to the physical qubit on the target
hardware. When applying a CNOT gate, the two qubits connected by
the CNOT gate need to be physically connected to each other. Due to the irregular
physical qubit layout of existing devices, it is generally considered impossible
to find an initial mapping that makes the entire circuit CNOT-compliant. The
common practice is to insert SWAP operations to remap the logical qubits. A swap operation exchanges the states of the
two input qubits of interest. 
As shown in Fig. \ref{fig:swapgate}, a SWAP
operation is implemented using 3 CNOT gates for architecture with
bi-directional links, or 3 CNOT gates plus 4 Hadamard gates for architecture
with single-direction links, where a bi-directional link means both ends of the
link can be the control or target qubit, while single-direction link means
only one end of it can be the control qubit.

IBM's Qiskit uses a stochastic method to insert SWAPs~\cite{IBMQiskit}
operations but often results in significant increase in the number of inserted
gates and depth. Existing works~\cite{Siraichi+:CGO18, Zulehner+:DATE18, Li+:ASPLOS19} are
more efficient than IBM's Qiskit mapper. They use efficient heuristics to find
the mapping rather than a stochastic method. However, the main objective of
these methods is to reduce the  gate count. It makes sense to
minimize the gate count, but it is more important to focus on the depth of
circuit, as in the NISQ era the depth is equivalent to the estimated execution
time. Reducing the depth of the circuit can reduce the likelihood of the circuit
failing at an early stage. 

\begin{figure}
\centering
\includegraphics[width=0.3\textwidth]{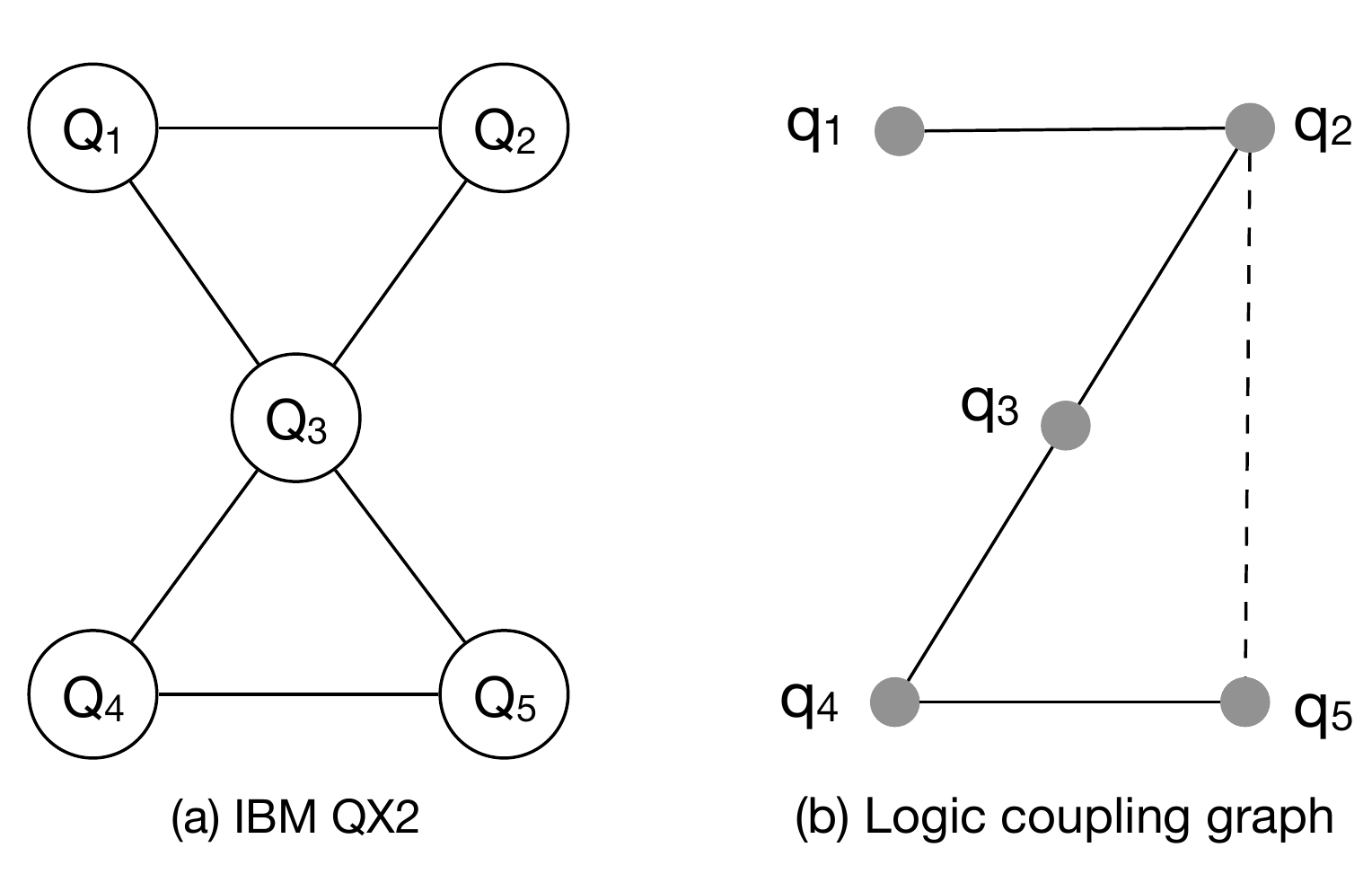}
\caption{(a) The connectivity structure of IBM QX2, (b) The coupling graph for
logical qubits in the motivation example in Fig. \ref{fig:motivation}}
\label{fig:hardware}
\end{figure}

\begin{figure}
\centering
\includegraphics[width=0.45\textwidth]{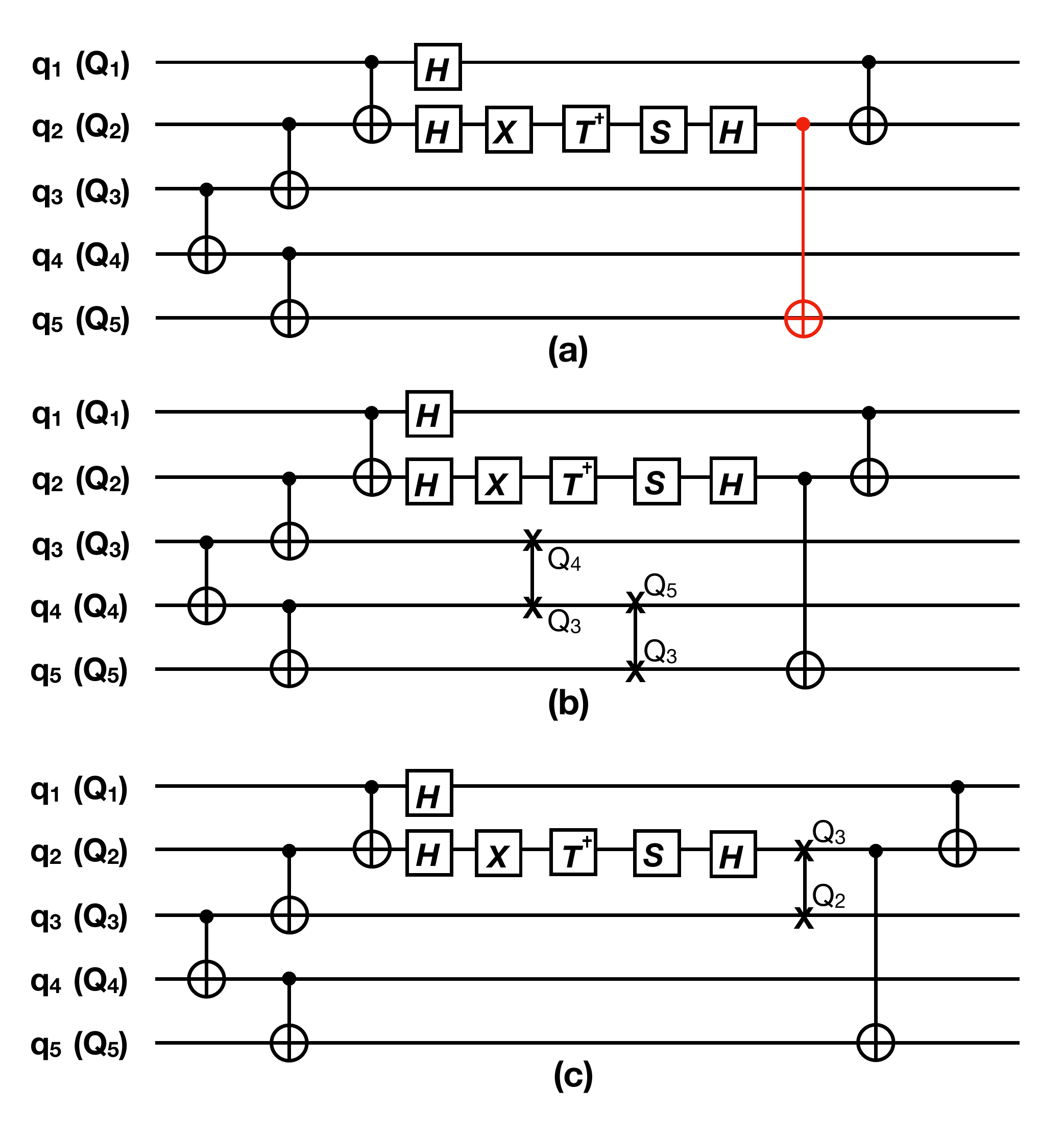}
\caption{Motivation Example: (a) the original logical circuit; (b) uses 2 swaps but the depth of the circuit is not increased; (c) only uses 1 swap but the depth of the circuit
has been increased }
\label{fig:motivation}
\end{figure}

We show an motivation example in Fig. \ref{fig:motivation}. The hardware model is shown in
Fig. \ref{fig:hardware} (a). It has five qubits and the connectivity is the same as the IBM
QX2 architecture except that the links are all bidirectional. There are 5 physical qubits: $Q_{1}$ to $Q_{5}$ and six
bi-directional edges. One CNOT gate can only be applied on one of these edges.

In the example, the initial mapping
between logical qubits (denoted by lower case \emph{q}) and physical qubits (denoted by
the upper case \emph{Q}) is shown next to each qubit (line), which is $\{$ $\{q_{1}
\rightarrow Q_{1}\}, \{q_{2}\rightarrow Q_{2}\}, \{q_{3}\rightarrow Q_{3}\},
\{q_{4}\rightarrow Q_{4}\}, \{q_{5}\rightarrow Q_{5}\}$ $\}$. With this initial
mapping, it starts scheduling gates one by one until it encounters a (set
of) CNOT gate(s) which cannot be scheduled due to physical constraints. We show the interaction of logical qubits in Fig. \ref{fig:hardware}(b) such that two logical qubits are connected if there is a CNOT operation between them. When we
encounter the gate ``CNOT $q_{2}$, $q_{5}$" (marked red in the circuit diagram in Fig. \ref{fig:motivation} and as the dotted line in the logical coupling graph Fig. \ref{fig:hardware}),
the scheduling has to terminate since this translates into ``CNOT $Q_{2}$,
$Q_{5}$" on the hardware, while no physical link exists between $Q_{2}$ and
$Q_{5}$. Necessary \emph{SWAP} operations are needed. When applying a SWAP
operation, the two input physical qubits will exchange their states. Fig.
\ref{fig:motivation} (b) and (c) provide two options for transforming the
circuit. Fig. \ref{fig:motivation} (b) inserts 2 SWAPs (\emph{SWAP $Q_3$, $Q_4$} and
\emph{SWAP $Q_3$, $Q_5$}) such that ``CNOT $q_{2}$, $q_{5}$" becomes ``CNOT $Q_{2}$,
$Q_{3}$", however the two SWAPs can run in parallel with existing single qubit gates in the
circuit, without having to increase the depth of the circuit. Fig. \ref{fig:motivation} (c) inserts only 1 SWAP (\emph{SWAP $Q_2$,
$Q_3$}) such that ``CNOT $q_{2}$, $q_{5}$" becomes ``CNOT $Q_{3}$,
$Q_{5}$", but it can not overlap with existing single-qubit gates in the circuit
and will only increase the depth of the circuit by 3 (assuming we use 3 gates to
implement the SWAP operation and each elementary gate takes 1 cycle in this example). 

In this example, the best two known approaches by Zulehner \etal
\cite{Zulehner+:DATE18} and Li \etal \cite{Li+:ASPLOS19} will both choose to
insert 1 SWAP since they only optimize the number of gates inserted into the
circuit (or the depth of the inserted gates), but not the depth of the
entire transformed circuit. This example stresses the importance of
depth-awareness in SWAP insertion schemes and motivates our work. 


\vspace{-1pt}
\section{Proposed Solution}
\vspace{-1pt}


\subsection{Metric}
As our work is a depth-aware SWAP insertion scheme, we first precisely define the metric for characterizing the depth of a circuit. In order to fully explain the metric, we need to introduce the concepts of \emph{dependency graph} and \emph{critical path}. 

The dependency graph represents the precedence relation between quantum gates in a logical quantum circuit. The definition is below:

\begin{definition}{Dependency Graph} :
The dependency graph of a quantum circuit $C$ with a set of gates $\Psi$ is a Directed Acyclic Graph $G_\psi=(\Psi, E_\psi)$, $E_\psi \subseteq \psi \times \psi$. A directed edge from node $\psi_1$ to node $\psi_2$ exists if and only if the output of gate $\psi_1$ is (part of) the input of gate $\psi_2$ in the quantum circuit $C$.
\end{definition}

The critical path is referred to as the longest path in the dependency graph. And the definition is below:

\begin{definition}{Critical Path} :
Given a dependency graph $G_\psi=(\Psi, E_\psi)$ of a quantum circuit. The critical path is $CP=Max(Path(\psi_1, \psi_2))$ $ s.t.$ $\psi_1, \psi_2 \in E_{\psi}$ $and$ $ \psi_1 \neq \psi_2$ 
\end{definition}

The depth is characterizing the number of execution steps of a quantum circuit, which is tantamount to the critical path length of the circuit. The longest path in the dependence graph describes the minimal number of steps the circuit needs in order for every gate's data dependence be resolved. In Algorithm \ref{algo}, we show how we calculate the critical path.
\vspace{-5pt}
\begin{algorithm}[htb]
\label{algo}
\SetAlgoLined
\SetKwInOut{Input}{Input}
\SetKwInOut{Output}{Output}
\Input{The circuit's dependency graph $G(V,E)$ }
\Output{The critical path $CP$}
\BlankLine


 earliest\_start = \{\}\;
 CP = 0\;

 \For{n $\in$ V in topological order}{
    temp = 0\;
    \For{p $\in$ V's predecessors} {
      \If {temp $<$ earliest\_start[p] + latency[p]}{
         temp = earliest\_start[p] + latency[p]\;
      } 
    }
    earliest\_start[n] = temp\;
      \If {CP $<$ temp + latency[n]} {
         CP = temp\;
      }
 }

return CP \;
 
 \caption{Calculate the Critical Path of a Circuit}
\end{algorithm}
\vspace{-5pt}

We first sort the nodes in the directed acyclic graph in topological order. Then we process the nodes in that order. For each node, we check the earliest start time for each of its predecessors, and add it by the latency of that predecessor, then we choose the maximum and use it as the earliest start time of this node. The maximum of all nodes' earliest start time added by their latency is the critical path length. 

We use the critical path length as the metric for ranking different swap insertion options.




\subsection{Framework Design}
With the metric precisely explained in previous section, now we continue to explain the work flow of our framework and the intuitions behind it.

Before delving into the details of this framework, we need to define the \emph{layer} and the \emph{coupling graph}.

\begin{figure}[htb]
  \centering
  \includegraphics[width=0.9\linewidth]{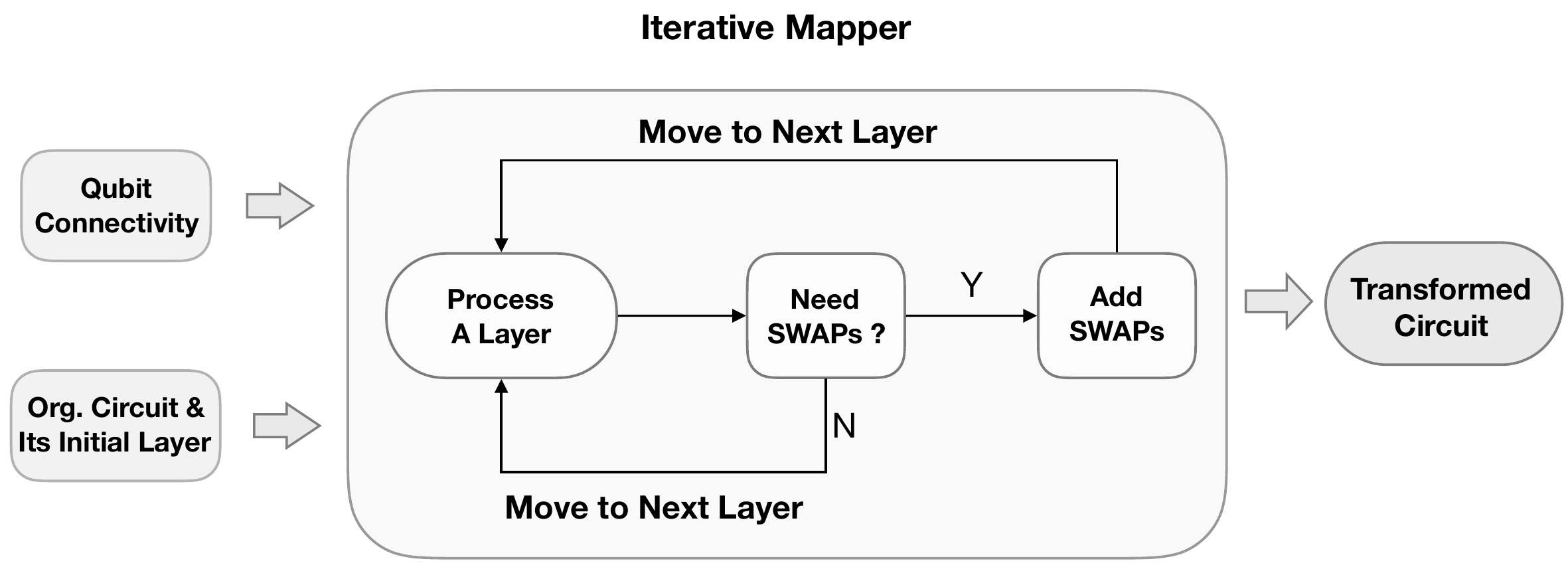}
  \caption{The Qubit Mapping Framework}
  \label{fig:workflow}
\end{figure}

\begin{definition}{Coupling Graph} : 
The coupling graph of a quantum architecture $X$ with a set of physical qubits $Q$ is a directed graph $G=(Q, E)$, $E\subseteq Q\times Q$. The edge $E_{x}=(Q_1,Q_2)\in E$ if and only if a $CNOT$ gate can be applied to $Q_1$ and $Q_2$ in $X$ with $Q_1$ being the control qubit and $Q_2$ being the target qubit. 
\end{definition}

We can divide the set of quantum gates in a circuit into layers, so that all gates in the same layer can be executed concurrently. The formal definition of a layer is:

\begin{definition}{Layer} :
A quantum circuit $C$ can be divided into layers $L = 
{l_1,l_2,l_3,...,l_m}$, while $\bigcup_{i=1}^{m}l_i = C$ and $\bigcap_{i=1}^{m}l_i = \emptyset$. The set of gates at layer $l_i$ can run concurrently  and  act on distinct sets of qubits.
\end{definition}
\vspace{-5pt}
To divide a circuit into layers, we group the gates that have the same \emph{earliest start time} (defined in Algorithm \ref{algo}) into the same layer. The order of the layers is thus determined by the order of the \emph{earliest start times}.   

We use an iterative process to find the mapping. Our framework is depicted in Fig. \ref{fig:workflow}. And this iterative process is explained as below.
We start the framework by taking the input of the coupling graph (also denoted as \emph{Qubit Connectivity}) and the original circuit's initial layer. 

We process the circuit layer by layer. Given a layer, we perform the following steps.
    \begin{itemize}
        \item We check the layer to see if it is hardware-compliant based on the coupling graph and the qubit mapping before current layer is scheduled.
        \item If YES, we move on to next layer.
        \item If NO, we invoke our mapping searcher to search for (the set of) swaps that are necessary to solve the current layer. We consider depth-awareness during the selection of the set of swap gates -- the resulted mapping of which generates the smallest critical path length (described in Section \ref{sec:map}). After we find a hardware-compliant mapping, we move to the next layer. 
    \end{itemize}
    
    After all layers are processed, the mapping terminates. 

\vspace{-1pt}
\subsection{Circuit Mapping Searcher}
\label{sec:map}
Here we describe the specific mapping searcher we use to overcome the coupling constraint for a given layer. 

We build our method upon the 
\emph{A-star} algorithm for finding valid mappings that minimize the number of only the inserted SWAP gates \cite{Zulehner+:DATE18}. We extend it by changing the ranking metric and allowing it to search for feasible mappings that do not necessarily have the smallest SWAP gate counts. It will help us search in a way that minimizes the depth while not significantly increasing the gate count.


We rank the swap options by the increase in the critical path length. Since it is an iterative process that handles the gates layer by layer, it is tempting to consider only minimizing the depth of the already processed circuit when deciding which swaps to use.  

\begin{figure}[htb]
  \centering
  \includegraphics[width=0.9\linewidth]{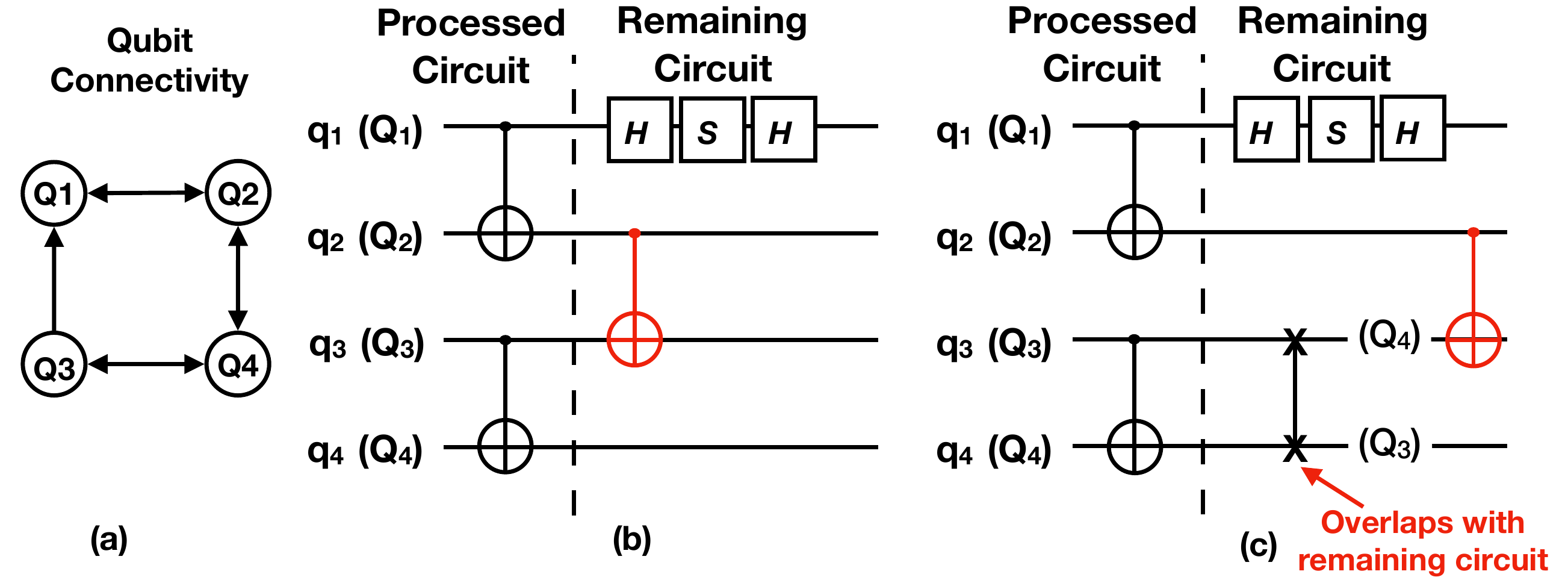}
  \caption{(a) Layout of an example architecture with 4 physical qubits (b) Example of a quantum circuit, the dashed line separates the processed circuit and the remaining circuit (c) Inserted SWAP overlaps with remaining circuit instead of existing processed circuit}
  \label{fig:remainingcircuit}
\end{figure} 

But the example in Fig. \ref{fig:remainingcircuit} shows that not only the processed circuit, but also the remaining circuit can help overlap the SWAPs with existing gates in the circuit without affecting the critical path.  As shown in Fig. \ref{fig:remainingcircuit}, for the CNOT gate (in red), there is no way it can overlap the necessary SWAPs with the processed circuit (dubbed as the circuit before the dashed line). But when we look after the dashed line, the three single-qubit gates can overlap with inserted SWAP. And this renders less impact to the depth of the resulting circuit, compared to if we insert the SWAP on $Q_1$ and $Q_2$.

\begin{figure}[!ht]
  \centering
  \includegraphics[width=1\linewidth]{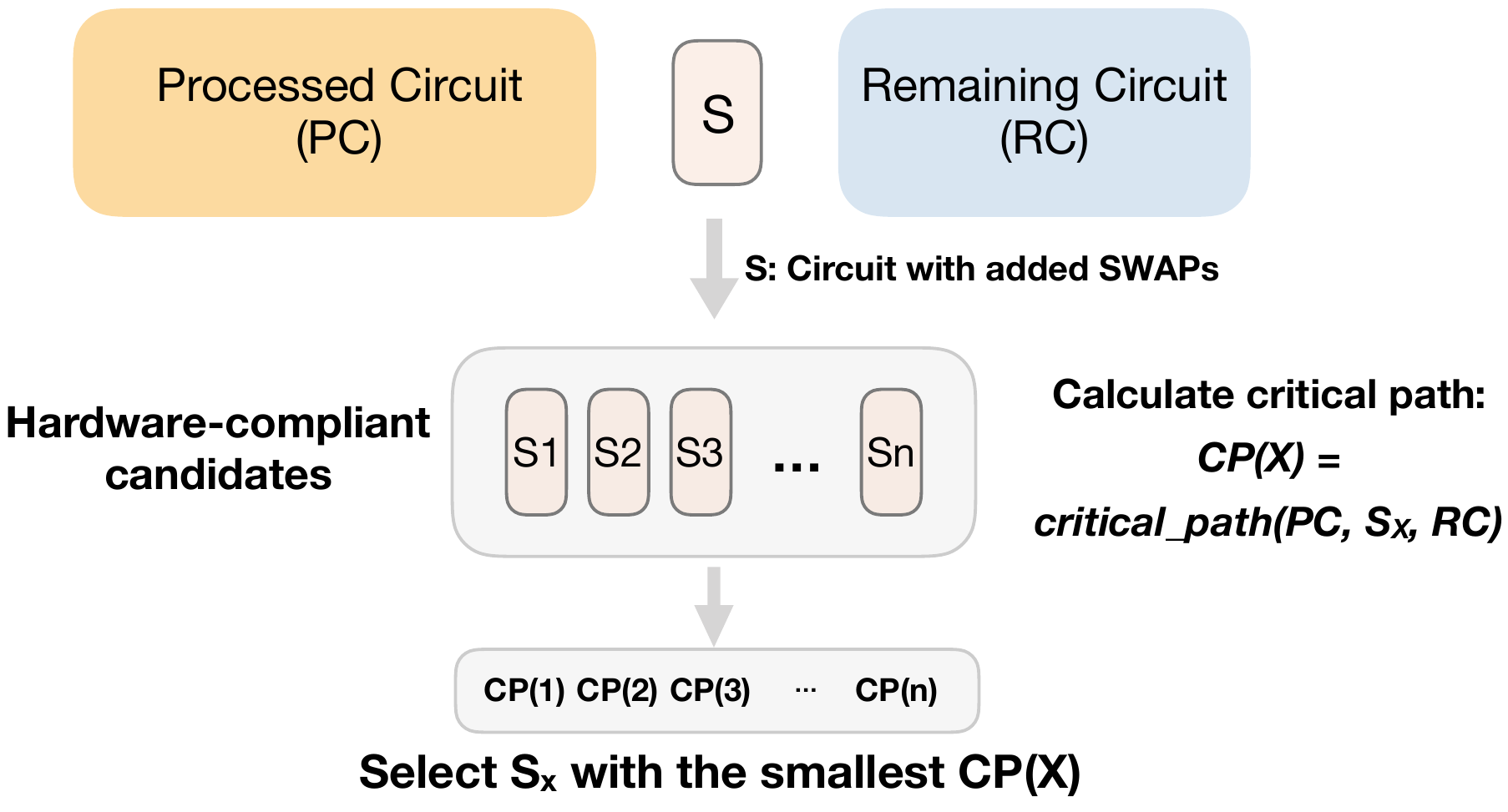}
  \caption{Choose SWAP Candidates}
  \label{fig:algo}
  \vspace{-5pt}
\end{figure}

Based on this intuition, we design our scheme of choosing the SWAP candidate as in Fig. \ref{fig:algo}. For each of the hardware-compliant remapping candidates that we acquire from the \emph{A-star} searcher, we calculate the critical path after merging the candidate (set of) swap(s) with both the processed circuit and the not-processed circuit. We choose the mapping that yields the shortest critical path.

\vspace{-5pt}
\subsection{Optimizations}
\vspace{-3pt}


We use two ways to optimize our proposed solution. One is to expand more nodes during the \emph{A-star} search, and another one is to search into deeper levels.
\subsubsection{Expand More Nodes}
In the search process for \emph{A-star}, the normal routine is to expand the one node of least cost at each step. Here, we can expand more than one node at each step and increase the search space. The number of nodes that can be expanded at a time can go from 1 to larger number. 
\subsubsection{Deeper Search}
 We increase the depth of the A-star search tree. In normal case, the search process ends when it finds the first node that minimizes the number of SWAPs, which is reflected as a certain level of the A-star tree. To this end, the second optimization that we applied here is to continue the search into a deeper level of the A-star tree. We can specify and tune the parameter of the deeper search. 

By tuning these parameters, there are more possible nodes added into our search space. With a larger search space, we have a larger possibility to jump out of one local optima and go to the global optima.

\section{Evaluation}
\label{sec:eval}
\vspace{-1pt}
\begin{table*}[!htb]
    \footnotesize
    \centering
    \caption{Summary of Experiment results}
    \label{tab:exp_table}
    \begin{tabular}{|l l|c||c|c|c||c|c|c|c||c|c|}
    \hline
         &  \multicolumn{2}{l||}{\textbf{Benchmark}}   & \multicolumn{3}{c||}{\textbf{Total Gate \#}} & \textbf{Depth} & \multicolumn{3}{c||}{\textbf{ Depth-delta}} & \multicolumn{2}{c|}{\textbf{Improvement}}\\
         \hline
         & name & n  & Zulehner  & Sabre & \textbf{DPS} & Original & Zulehner & Sabre &
\textbf{DPS} & Min & Max \\ \hline
         & 4gt5\_75 & 5 & 131 & 122 & 119 & 47 & 44 & 44 & 29 & 1.52 &	1.52 \\ \hline
         & mini-alu\_167 & 5 &	435	&396 &432	&162 & 131 & 125 & 119 & 1.05 &	1.10  \\ \hline
         & mod10\_171 & 5 & 361 & 328 & 298 & 139 & 117 & 89 & 39 & 2.28 &	3  \\ \hline
         & alu-v2\_30 & 6 &	804	&717&795&		285&	261&	241&	201&	1.20 &	1.30  \\ \hline
         & decod24-enable\_126 &	6 & 533&	476&509&		190&	187&	150&	141&	1.06&	1.33  \\ \hline
         & mod5adder\_127&	6 & 849	&780&858&		302&	256&	256&	222	&1.15&	1.15  \\ \hline
         & 4mod5-bdd\_287&	7 & 94&	94&94&		41&	18&	23&	17&	1.06&	1.35  \\ \hline
         & alu-bdd\_288&	7 & 126&	117&135&		48&	36&	36&	30&	1.2&	1.2  \\ \hline
         & majority\_239&7	&915&	780&885&		344&	265&	194&	182&	1.06&	1.46  \\ \hline
         & rd53\_130&	7&	1619&	1508&1619&		569&	529&	482&	384&	1.26&	1.38  \\ \hline
         & rd53\_135&	7&	419&	410&422&		159&	116&	112&	109&	1.03&	1.06  \\ \hline
         & rd53\_138&	8&	186&	183&174&		56&	37&	40&	21&	1.76&	1.90  \\ \hline
         & cm82a\_208&	8&	899&	944&1007&		337&	219&	295&	213&	1.03&	1.38  \\ \hline
         & qft\_10	&10	&266	&263&281		&63	&47	&96	&44	&1.07	&2.18  \\ \hline
         & rd73\_140&	10&	347&	329&338&		92&	84&	79&	67&	1.18	&1.25  \\ \hline
         & dc1\_220&	11&	2868&	2685&3129&		1038&	820&	697&	681&	1.02&	1.20  \\ \hline
         & wim\_266	&11&	1505&	1415&	1511&	514&	431&	450&	311	&1.39&	1.45  \\ \hline
         & z4\_268	&11	&4453&	4477&4972&		1644&	1162&	1492&	1076&	1.08&	1.39  \\ \hline
         & cycle10\_2\_110&	12&	9143&	8666&10115&		3386&	2467&	2640&	2421&	1.02&	1.09  \\ \hline
         & sym9\_146	&12&	493&	454&472&		127&	118&	138&	86&	1.37&	1.60 \\ \hline
         & adr4\_197&	13&	5299&	5017&5530&		1839&	1439&	1599&	1210&	1.19&	1.32 \\ \hline
         & rd53\_311&	13&	467&	413&446&		124&	138&	157&	87&	1.59&	1.80 \\ \hline
         & cnt3-5\_179	&16&	325&	238&286&		61&	79&	59&	43&	1.37&	1.84 \\ \hline
         
    \end{tabular}
    \vspace{2ex}

     {\raggedright We compare the total gate count generated.
For depth, we compare the increased depth for each benchmark, denoted as
``Depth-delta" here. The improvement represents the ratio of a baseline's
depth-delta divided by DPS's depth-delta. Min/Max represents the improvement over the best/worst baseline. \par}

\end{table*}

\begin{figure}[!ht]
  \centering
  \includegraphics[width=0.3\linewidth]{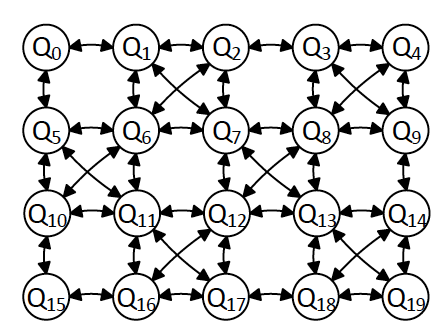}
  \caption{IBM Q20 Tokyo Physical Layout \cite{Li+:ASPLOS19} }
  \label{fig:ibmq20tokyo}
\end{figure}

In this section, we evaluate our \textbf{d}e\textbf{p}th-aware \textbf{s}wap insertion scheme
(denoted as DPS) and compare
it with the two state-of-the-art qubit mappers. The experiment setup is listed below:
\begin{itemize}
    \item \textbf{Benchmarks}: We use the quantum circuits from
RevLib~\cite{Wille+:ISMVL08}, IBM Qiskit~\cite{IBMQiskit}, and ScaffCC~\cite{JavadiAbhari+:CCF14}. 
    \item \textbf{Hardware Model}: We use IBM's 20-qubit Q20 Tokyo architecture, which
was used in \cite{Li+:ASPLOS19}'s work. The
qubit connectivity graph is shown in Fig. \ref{fig:ibmq20tokyo}.
    \item \textbf{Evaluation Platform}: The mapping experiments are conducted on a Intel
2.4 GHz Core i5 machine, with 8 GB 1600 MHz DDR3 memory. The operating system is
MacOS Mojave. We use IBM's Qiskit \cite{IBMQiskit} to evaluate the depth of the transformed
circuit. 
    \item \textbf{Baselines}: We compare our work with two best know qubit
mapping solutions, the work by Zulehner and others ~\cite{Zulehner+:DATE18}
(denoted as \emph{Zulehner}), the Sabre qubit mapper from~\cite{Li+:ASPLOS19}
(denoted as \emph{Sabre}), and
IBM's stochastic mapper in Qiskit. Since IBM's Qiskit mapper is significantly
worse in terms of gate count and depth than all other mappers we evaluate, as also evidenced in the work by Zulehner
\etal \cite{Zulehner+:DATE18}, we do not present Qiskit results. 
    \item \textbf{Metrics}: We are comparing the depth and gate count of the
transformed circuit circuits for all different strategies. 
\end{itemize}

 Table. \ref{tab:exp_table} shows a summary experimental results. For gate
count, we compare the total gate count generated in the transformed circuit. For
depth, we compare the increased depth for each benchmark, denoted as
``Depth-delta" in Table \ref{tab:exp_table}. The improvement columns provides
the ratio between one of the two baseline's \emph{depth-delta} and our
\emph{depth-delta}. We use the term \emph{minimum improvement} to denote the improvement over
the best of the two baselines, and the term \emph{maximum improvement} to denote
the improvement over the worse of the two baselines.
 
 We discuss our findings from the following three aspects: depth reduction, gate
count change, and the trade-off between gate count and depth.  

\vspace{-5pt}
\subsection{Depth Reduction}
\vspace{-5pt}
For depth reduction, as shown in Table \ref{tab:exp_table}, our proposed
solution outperforms the two baselines
\emph{Zulehner} and \emph{Sabre}. Comparing \emph{depth-delta}, the added depth
of the circuit, our approach outperforms the
better of the two baselines by more than 20\% and up to 3X. For five out of the twenty-three
benchmarks, our improvement on \emph{depth-delta} is less than 20\% compared
with the better of the two baselines. However, for these cases, our approach
still achieves considerable improvement over the worse of the two baselines. In
these cases, it is possible that one of the two baselines happen to achieve very
good depth in the transformed circuit and there is not much potential to
improve. But our approach is still able to find a good mapping for
these benchmarks and the performance is on par with the better of the two
baselines. 

\vspace{-5pt}
\subsection{Gates Count Changes}
\vspace{-5pt}
The primary goal of our depth-aware qubit mapper is to minimize the depth of the
circuit. However, we discover that our qubit mapper can sometimes reduce the gate
count. We discover that four out of the twenty three
(17\%) benchmarks, our qubit mapper yields the smallest number of gates
among all three versions of qubit mappers. For 57\% of these benchmarks, our
method is ranked among top-2 of the three qubit mappers in terms of gate count. For
the benchmarks where our method yields the largest gate count, the increased
gate count percentage is negligible. On average, our depth-aware qubit mapper
adds 3\% gate count. From the experiment results, we can see that our solution
does not greatly increase the number of gates while reducing the depth of the
circuit.

\vspace{-5pt}
\subsection{Trade-off between Gate Count and Depth}
\vspace{-5pt}
While all previous works focus on reducing the total gate count (and the
depth among the inserted gates themselves) after qubit mapping transformation,
it is crucial to think about the trade-off between the resulted gate count and
depth. Sometimes the choice made during the search process that favors the
reduced gate count, might adversely affect the critical path. In Table  \ref{tab:exp_table}, the \emph{Sabre} mapper reduces the number of gates
for 10-qubit QFT by 1.1\% compared with \emph{Zulehner}'s mapper, but increases the
depth by 44.5\%. For the \textit{sym\_9\_246} benchmark, \emph{Sabre} reduces
the gate count by 3.8\% compared with our approach, but increases the depth by
25.5\%. Therefore a small reduction in the gate count may not be worthwhile if
it increases the circuit depth significantly.

\section{conclusion}
\vspace{-5pt}
The physical layout of contemporary quantum devices imposes limitations for
mapping a high level quantum program to the hardware. It is critical to develop
an efficient qubit mapper in the NISQ era. Existing
studies aim to reduce the gate count but are oblivious to the depth of the
transformed circuit. This paper presents the design of the
first depth-aware swap insertion scheme. Experiment results show that our
proposed solution generates hardware-compliant circuits with reduced depth
compared with state-of-the-art mapping schemes, with negligible overhead of increased gate count.


\vspace{-8pt}
\bibliographystyle{IEEEtran}
\bibliography{main}

\end{document}